\documentstyle[12pt,epsf,feynarts]{article}

\newcommand\pubdate{\today}

\newcommand\hepnumber{hep-ph/0212273}

\def\csumb{Ottawa-Carleton Institute for Physics,
Department of Physics, Carleton University, Ottawa, Canada K1S 5B6
}

\def\Title#1{\begin{center} {\Large\bf #1 } \end{center}}
\def\Author#1{\begin{center}{ \sc #1} \end{center}}
\def\Address#1{\begin{center}{ \it #1} \end{center}}

\newcommand\pubblock{\rightline{\begin{tabular}{l}
         \pubdate\\ \hepnumber \end{tabular}}}
\newenvironment{Abstract}{\begin{quotation}  }{\end{quotation}}

\def\beq{\begin{equation}}
\def\eeq{\end{equation}}
\def\beqn{\begin{eqnarray}}
\def\eeqn{\end{eqnarray}}
\def\bea{\begin{eqnarray}}
\def\eea{\end{eqnarray}}
\def\be{\begin{equation}}
\def\ee{\end{equation}}

\begin{document}
\begin{titlepage}
\pubblock

\Title{
Supersymmetric QCD corrections to lightest Higgs boson
associated production with top quark pair at Linear Colliders
}
\vfill
\Author{ Shouhua Zhu \footnote{E-mail address: huald@physics.carleton.ca}  }
\Address{\csumb}
\vfill
\begin{Abstract}
Supersymmetric (SUSY) QCD corrections to the lightest neutral
Higgs boson associated production with top quark pair are studied
in the minimal supersymmetric standard model (MSSM) at Linear
colliders. Our calculations show that the SUSY QCD effects
generally are very moderate (say 10\%) and under control, except
for some rescattering effects which lead to a breakdown of
perturbation theory and require a more detailed study. In the
vicinity of the production threshold for the favorable model
parameters under the framework of the on-shell renormalization
scheme, SUSY QCD can be as large as about -50\%. Such effects
might be acted as the probe to determine the sign of $M_{LR}\equiv
A_t-\mu/\tan\beta$.

\end{Abstract}
\vfill


\end{titlepage}

\eject \baselineskip=0.3in


\section{Introduction}


One of the key issues of present high energy physics is
to check whether the Higgs field(s) account for the mechanism of the mass generation.
In the standard model
(SM) of the high energy physics, only one physical Higgs boson exists after
symmetry spontaneously broken, while
five Higgs bosons in the minimal supersymmetric standard model (MSSM).

Until now, the precise high energy experiments have tested the
gauge bosons self interactions and the gauge bosons interactions
with fermions (except top sector) of the SM with very good
accuracy, while the Higgs sector is totally untouched
\cite{smtest}. In the SM and MSSM, the couplings among Higgs
bosons and top quarks play a very unique role because which are
not suppressed by $\frac{m_q}{m_W}$. Very likely, these couplings
are firstly detected at experiments. Besides the aspect of the
study on the properties of the Higgs bosons, there is another
strong motivation to examine the $H-t-\bar t$ coupling: searching
new physics beyond the SM. It is commonly thought that new physics
might appear in the top sector for its huge mass.

Supersymmetric (SUSY) QCD interactions by exchanging gluinos and
squarks might impact greatly on the Higgs-top quarks vertexes. If
the Higgs bosons are heavy enough, the SUSY QCD effects can be
studied in the decay channel $H \rightarrow t \bar t$ (H is
general notation for Higgs boson) \cite{higgsstudy}. Otherwise,
production channels offer alternative methods and should be
carefully examined. At hadron colliders, Higgs boson can be
associated production with top quark pair through $gg$ and/or
$q\bar q$ sub-processes \cite{ggqqtth}. Besides hadron colliders,
as a supplementary machine, the linear colliders (LC) will provide
ideal place to study such effects. Even in the seventies, the
process $e^+e^- \rightarrow H \bar t t$ in the SM at LC has been
proposed, which can be used as finding Higgs boson and measuring
the couplings of $g_{H\bar t t}$
 \cite{eetthtree}.
The calculations of the important QCD effects for the process are
not accomplished until the end of the last century \cite{eetthqcd}.
In the MSSM, the neutral Higgs boson associated production with top quark pair
at LC including the QCD effect are available soon after \cite{eettsusyhqcd}. However, the SUSY QCD effects
on the process $e^+ e^- \rightarrow t \bar t H$
are not touched, which is the subject of this paper. In this paper, we focus on
the lightest neutral Higgs boson $h^0$ associated production with top quark pair in the MSSM.

\section{Basics and technical set-ups}
%

The diagrams including the SUSY QCD corrections are shown in Fig.1-3.
These diagrams and Feynman amplitudes are generated by FeynArts \cite{fa}.
Fig. 1 and 2 are the Feynman diagrams at tree and one-loop level, respectively.
In Fig. 3, we draw the diagrams containing counter-terms.
 In this paper, we renormalize the quark wave function and mass in
 the on-shell (OS) renormaliztion
scheme. Therefore in Fig. 3, we do not show the diagrams with counter-terms at external
legs because
 in the OS renormaliztion
scheme, such kind of diagrams are completely eliminated by the corresponding virtual diagrams.

The analytical results for Feynman amplitudes and phase space integrations are tedious
but easily obtained through standard procedure, which are useless to almost all
readers. Therefore, we decide to omit the explicit analytical results here
\footnote{Programs are available on request.}.
For the calculations of the amplitudes and the Monte Carlo phase
 space integrations, we rely heavily on the popular packages FormCalc and
 LoopTools \cite{fclt} \footnote{A technical detail, LoopTools is not applicable for some $D_0$ with certain parameters that all $(m_i^2+m_j^2-p_{ij})^2 <4 m_i^2 m_j^2$, details of the definition and convention for $D_0$ can be found in manual of LoopTools or in Ref \cite{Dennerhbs}.
 For this exceptional case, the new program has been added according to Ref \cite{Dennerhbs}.}.
Before we shift to numerical discussions, it should be emphasized that
the entire calculations and the cancellation of ultraviolet divergences have been
checked carefully.

\section{Numerical discussions and summary}

In the MSSM, there are dozens of parameters and experiments
constrains, for examples $\Delta \rho$ and direct mass limits for
gluino, squarks and Higgs bosons. We rely on the built-in programs
of FormCalc to calculate the SUSY particles mass spectrum (we use
two-loop results for Higgs mass calculations) and SUSY parameters
as well as set constraints on known limits from experiments.
Specifically, we set
\begin{eqnarray}
\Delta \rho < 3\times 10^{-3}
\end{eqnarray}
\begin{eqnarray}
m_{\tilde{g}} >175 \ GeV, \ \ \ &
m_{\tilde{t}}>80 \ GeV, \ \ \ & m_{\tilde{b}}> 70 \ GeV
\nonumber \\
m_{\tilde{q}}>150 GeV, \ \ \  & m_{h^0}>91 GeV. \ \ \ &
\end{eqnarray}
For simplicity we have assumed the SUSY parameters $M_Q=M_U=M_D=M_S$ and
$A_t=A_b$.

As for the first attempt, we would like to see the SUSY QCD effects for the non-mixing case,
i.e. the non-diagonal matrix elements $M_{LR} \equiv A_t-\mu/\tan\beta$ for stop mass matrix
are zero. Our numerical results show that the SUSY QCD effects are small for such case,
and the relative corrections $\delta=(\sigma_{NLO}-\sigma_{LO})/\sigma_{LO}$ are at most
about two percents, which are hard to detect at experiments.
In the following we will focus on the mixing case.

After scanning the whole parameters space, in order to demonstrate
the numerical results, in Fig. 4-6 we show the cross sections and
relative corrections $\delta$ for a specific parameter set as a
function of $\tan\beta$, center-of-mass energy $\sqrt{s}$ of
$e^+e^-$, as well as the mass of gluino. From the figures, we can
see that the next-to-leading order (NLO) SUSY QCD effects decrease
the LO cross sections and the magnitude can reach $50\%$. For such
kind of large corrections, it should be noted that main
contributions come from diagrams (5)-(8) of Fig. 2, and the
lighter stop mass is about 100 GeV in this case. Similar to the SM
QCD effects, the large corrections occur in the vicinity of the
production threshold (Fig. 5), where the cross sections become
small. Furthermore, from Fig. 6, it is obviously that the
corrections are at the peak around $m_{\tilde{g}}=190 \sim 200$
GeV, where the re-scattering enhancement $t^* \rightarrow
\tilde{g}+ \tilde{t}_1 \rightarrow t + h^0$ appears  at
$m_{\tilde{g}}+ m_{\tilde{t}_1} \sim m_t+m_{h^0}$.

In table 1, we show that cross sections at LO and NLO as well as
the relative corrections as a function of $A_t$. The large SUSY
QCD effects emerge only with positive large $A_t$ [or $M_{LR}$],
which ensures that the large contributions from lighter
stops don't vanish.

From our parameters space scanning, in order to have the large
SUSY QCD effects, the following requirements should be satisfied
\begin{enumerate}
\item there must be large mass splitting between stops, moreover
$M_{LR}$ should be large and positive,
\item $M_S$ should not be large,
\item gluino mass should not be heavy.
\end{enumerate}
As mentioned above, condition (1) prevents the cancellation
between stops, and makes one stop mass light which is essential
for large SUSY QCD effects, especially in the threshold regions.
Conditions (2) and (3) ensure that SUSY effects do not decouple from
the process. \\

To summarize, the SUSY QCD corrections to the lightest neutral
Higgs boson associated production with top quark pair are studied
in the MSSM at LC. Our studies show that generally the corrections
are very moderate (say 10\%) and under control, except for some
rescattering effects which lead to a breakdown of perturbation
theory and require a more detailed study, resummation or something
of this kind. Contrary to the QCD corrections arising from gluons
which are about 50\% for $m_{h^0}=120$ GeV and $\sqrt{s}=500$ GeV
[near the production threshold], the SUSY QCD effects can be also
as large as about -50\% in this kinemical region for the favorable
model parameter space under OS renormalization scheme. Such kind
of large SUSY QCD effects might be the probe to determine the sign
of $M_{LR}$. We note that if nature does choose lighter stops,
which can be copiously produced at hadron colliders and linear
colliders. Comparison between direct productions and indirect
virtual effects might give more information on the MSSM.

\section*{Acknowledgements}
The author would like to thank Dr. Thomas Hahn, Dr. Christian Schappacher
and Dr. G.J. van Oldenborgh
for the help on the packages used in this paper.
This work was supported in part by the Nature Sciences and Engineering Research Council of Canada, the Alexander von Humboldt
Foundation, National Nature Science Foundation of China.\\



\newpage

\begin{figure}
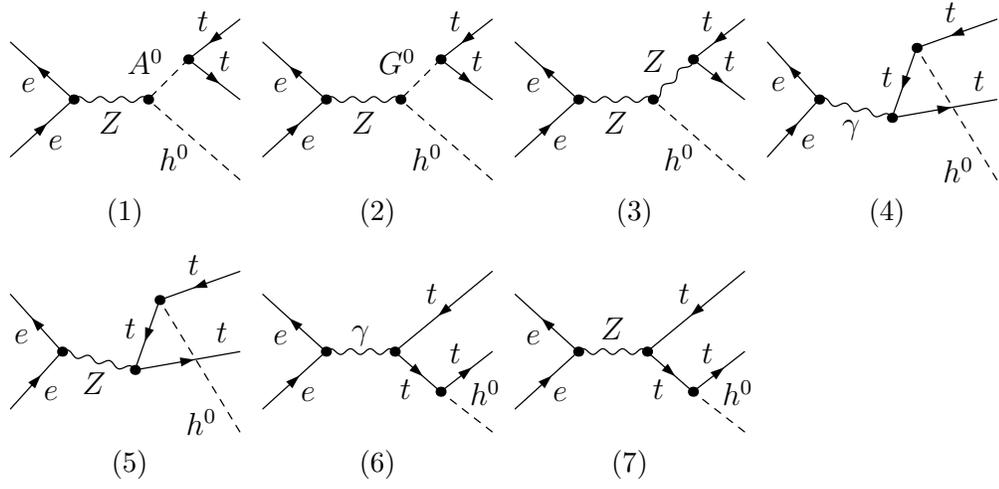

\unitlength=1bp%
\begin{feynartspicture}(432,504)(4,5.3)

\FADiagram{(1)}
\FAProp(0.,15.)(5.5,10.)(0.,){/Straight}{-1}
\FALabel(2.18736,11.8331)[tr]{$e$}
\FAProp(0.,5.)(5.5,10.)(0.,){/Straight}{1}
\FALabel(3.31264,6.83309)[tl]{$e$}
\FAProp(20.,17.)(15.5,13.5)(0.,){/Straight}{1}
\FALabel(17.2784,15.9935)[br]{$t$}
\FAProp(20.,10.)(15.5,13.5)(0.,){/Straight}{-1}
\FALabel(18.2216,12.4935)[bl]{$t$}
\FAProp(20.,3.)(12.,10.)(0.,){/ScalarDash}{0}
\FALabel(15.6239,6.00165)[tr]{$h^0$}
\FAProp(5.5,10.)(12.,10.)(0.,){/Sine}{0}
\FALabel(8.75,8.93)[t]{$Z$}
\FAProp(15.5,13.5)(12.,10.)(0.,){/ScalarDash}{0}
\FALabel(13.3108,12.1892)[br]{$A^0$}
\FAVert(5.5,10.){0}
\FAVert(15.5,13.5){0}
\FAVert(12.,10.){0}

\FADiagram{(2)}
\FAProp(0.,15.)(5.5,10.)(0.,){/Straight}{-1}
\FALabel(2.18736,11.8331)[tr]{$e$}
\FAProp(0.,5.)(5.5,10.)(0.,){/Straight}{1}
\FALabel(3.31264,6.83309)[tl]{$e$}
\FAProp(20.,17.)(15.5,13.5)(0.,){/Straight}{1}
\FALabel(17.2784,15.9935)[br]{$t$}
\FAProp(20.,10.)(15.5,13.5)(0.,){/Straight}{-1}
\FALabel(18.2216,12.4935)[bl]{$t$}
\FAProp(20.,3.)(12.,10.)(0.,){/ScalarDash}{0}
\FALabel(15.6239,6.00165)[tr]{$h^0$}
\FAProp(5.5,10.)(12.,10.)(0.,){/Sine}{0}
\FALabel(8.75,8.93)[t]{$Z$}
\FAProp(15.5,13.5)(12.,10.)(0.,){/ScalarDash}{0}
\FALabel(13.3108,12.1892)[br]{$G^0$}
\FAVert(5.5,10.){0}
\FAVert(15.5,13.5){0}
\FAVert(12.,10.){0}

\FADiagram{ (3)}
\FAProp(0.,15.)(5.5,10.)(0.,){/Straight}{-1}
\FALabel(2.18736,11.8331)[tr]{$e$}
\FAProp(0.,5.)(5.5,10.)(0.,){/Straight}{1}
\FALabel(3.31264,6.83309)[tl]{$e$}
\FAProp(20.,17.)(15.5,13.5)(0.,){/Straight}{1}
\FALabel(17.2784,15.9935)[br]{$t$}
\FAProp(20.,10.)(15.5,13.5)(0.,){/Straight}{-1}
\FALabel(18.2216,12.4935)[bl]{$t$}
\FAProp(20.,3.)(12.,10.)(0.,){/ScalarDash}{0}
\FALabel(15.6239,6.00165)[tr]{$h^0$}
\FAProp(5.5,10.)(12.,10.)(0.,){/Sine}{0}
\FALabel(8.75,8.93)[t]{$Z$}
\FAProp(15.5,13.5)(12.,10.)(0.,){/Sine}{0}
\FALabel(13.134,12.366)[br]{$Z$}
\FAVert(5.5,10.){0}
\FAVert(15.5,13.5){0}
\FAVert(12.,10.){0}

\FADiagram{ (4)}
\FAProp(0.,15.)(4.5,10.)(0.,){/Straight}{-1}
\FALabel(1.57789,11.9431)[tr]{$e$}
\FAProp(0.,5.)(4.5,10.)(0.,){/Straight}{1}
\FALabel(2.92211,6.9431)[tl]{$e$}
\FAProp(20.,17.)(13.,14.5)(0.,){/Straight}{1}
\FALabel(15.9787,16.7297)[b]{$t$}
\FAProp(20.,10.)(10.85,8.4)(0.,){/Straight}{-1}
\FALabel(18.4569,10.7663)[b]{$t$}
\FAProp(20.,3.)(13.,14.5)(0.,){/ScalarDash}{0}
\FALabel(17.98,4.93)[tr]{$h^0$}
\FAProp(4.5,10.)(10.85,8.4)(0.,){/Sine}{0}
\FALabel(7.29629,8.17698)[t]{$\gamma$}
\FAProp(13.,14.5)(10.85,8.4)(0.,){/Straight}{1}
\FALabel(10.9431,11.9652)[r]{$t$}
\FAVert(4.5,10.){0}
\FAVert(13.,14.5){0}
\FAVert(10.85,8.4){0}

\FADiagram{ (5)}
\FAProp(0.,15.)(4.5,10.)(0.,){/Straight}{-1}
\FALabel(1.57789,11.9431)[tr]{$e$}
\FAProp(0.,5.)(4.5,10.)(0.,){/Straight}{1}
\FALabel(2.92211,6.9431)[tl]{$e$}
\FAProp(20.,17.)(13.,14.5)(0.,){/Straight}{1}
\FALabel(15.9787,16.7297)[b]{$t$}
\FAProp(20.,10.)(10.85,8.4)(0.,){/Straight}{-1}
\FALabel(18.4569,10.7663)[b]{$t$}
\FAProp(20.,3.)(13.,14.5)(0.,){/ScalarDash}{0}
\FALabel(17.98,4.93)[tr]{$h^0$}
\FAProp(4.5,10.)(10.85,8.4)(0.,){/Sine}{0}
\FALabel(7.29629,8.17698)[t]{$Z$}
\FAProp(13.,14.5)(10.85,8.4)(0.,){/Straight}{1}
\FALabel(10.9431,11.9652)[r]{$t$}
\FAVert(4.5,10.){0}
\FAVert(13.,14.5){0}
\FAVert(10.85,8.4){0}

\FADiagram{(6)}
\FAProp(0.,15.)(5.5,10.)(0.,){/Straight}{-1}
\FALabel(2.18736,11.8331)[tr]{$e$}
\FAProp(0.,5.)(5.5,10.)(0.,){/Straight}{1}
\FALabel(3.31264,6.83309)[tl]{$e$}
\FAProp(20.,17.)(11.5,10.)(0.,){/Straight}{1}
\FALabel(15.2447,14.2165)[br]{$t$}
\FAProp(20.,10.)(15.5,6.5)(0.,){/Straight}{-1}
\FALabel(17.2784,8.9935)[br]{$t$}
\FAProp(20.,3.)(15.5,6.5)(0.,){/ScalarDash}{0}
\FALabel(18.0681,5.29616)[bl]{$h^0$}
\FAProp(5.5,10.)(11.5,10.)(0.,){/Sine}{0}
\FALabel(8.5,11.07)[b]{$\gamma$}
\FAProp(11.5,10.)(15.5,6.5)(0.,){/Straight}{1}
\FALabel(12.9593,7.56351)[tr]{$t$}
\FAVert(5.5,10.){0}
\FAVert(11.5,10.){0}
\FAVert(15.5,6.5){0}

\FADiagram{(7)}
\FAProp(0.,15.)(5.5,10.)(0.,){/Straight}{-1}
\FALabel(2.18736,11.8331)[tr]{$e$}
\FAProp(0.,5.)(5.5,10.)(0.,){/Straight}{1}
\FALabel(3.31264,6.83309)[tl]{$e$}
\FAProp(20.,17.)(11.5,10.)(0.,){/Straight}{1}
\FALabel(15.2447,14.2165)[br]{$t$}
\FAProp(20.,10.)(15.5,6.5)(0.,){/Straight}{-1}
\FALabel(17.2784,8.9935)[br]{$t$}
\FAProp(20.,3.)(15.5,6.5)(0.,){/ScalarDash}{0}
\FALabel(18.0681,5.29616)[bl]{$h^0$}
\FAProp(5.5,10.)(11.5,10.)(0.,){/Sine}{0}
\FALabel(8.5,11.07)[b]{$Z$}
\FAProp(11.5,10.)(15.5,6.5)(0.,){/Straight}{1}
\FALabel(12.9593,7.56351)[tr]{$t$}
\FAVert(5.5,10.){0}
\FAVert(11.5,10.){0}
\FAVert(15.5,6.5){0}

\end{feynartspicture}
\caption[]{ Born diagrams of $e^+e^- \rightarrow t \bar t h^0$.}
\end{figure}

\begin{figure}
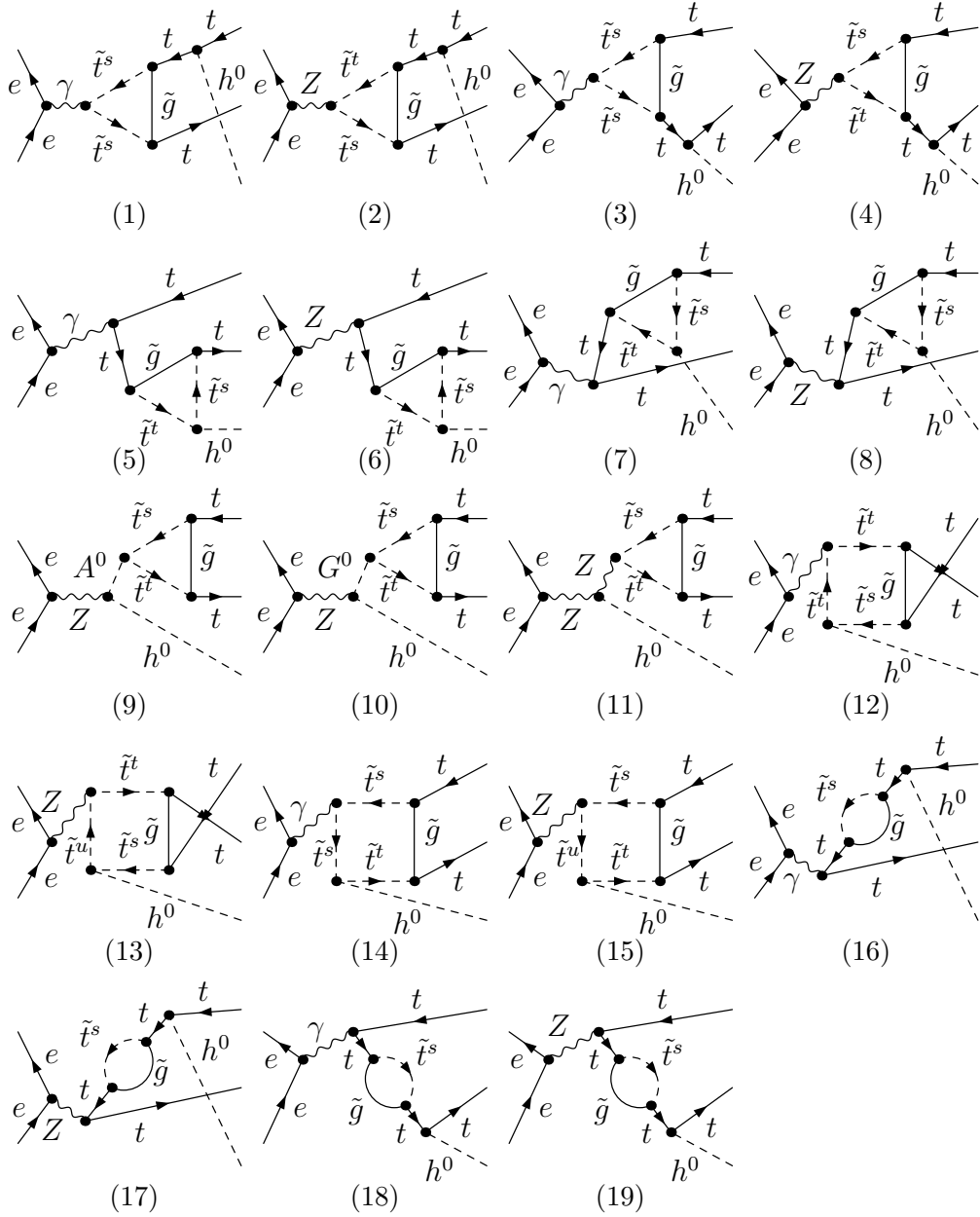


\unitlength=1bp%

\begin{feynartspicture}(432,504)(4,5.3)

\FADiagram{(1)}
\FAProp(0.,15.)(2.5,10.)(0.,){/Straight}{-1}
\FALabel(0.343638,12.2868)[tr]{$e$}
\FAProp(0.,5.)(2.5,10.)(0.,){/Straight}{1}
\FALabel(2.15636,7.28682)[tl]{$e$}
\FAProp(20.,17.)(16.,15.)(0.,){/Straight}{1}
\FALabel(17.7868,16.9064)[br]{$t$}
\FAProp(20.,10.)(12.,6.5)(0.,){/Straight}{-1}
\FALabel(14.6702,6.55096)[tl]{$t$}
\FAProp(20.,3.)(16.,15.)(0.,){/ScalarDash}{0}
\FALabel(17.87,12.55)[l]{$h^0$}
\FAProp(2.5,10.)(6.,10.)(0.,){/Sine}{0}
\FALabel(4.25,11.07)[b]{$\gamma$}
\FAProp(16.,15.)(12.,13.5)(0.,){/Straight}{1}
\FALabel(13.4558,15.2213)[b]{$t$}
\FAProp(12.,6.5)(6.,10.)(0.,){/ScalarDash}{-1}
\FALabel(8.699,7.39114)[tr]{$\tilde t^s$}
\FAProp(12.,6.5)(12.,13.5)(0.,){/Straight}{0}
\FALabel(12.82,10.)[l]{$\tilde g$}
\FAProp(6.,10.)(12.,13.5)(0.,){/ScalarDash}{-1}
\FALabel(8.699,12.6089)[br]{$\tilde t^s$}
\FAVert(2.5,10.){0}
\FAVert(16.,15.){0}
\FAVert(12.,6.5){0}
\FAVert(6.,10.){0}
\FAVert(12.,13.5){0}

\FADiagram{(2)}
\FAProp(0.,15.)(2.5,10.)(0.,){/Straight}{-1}
\FALabel(0.343638,12.2868)[tr]{$e$}
\FAProp(0.,5.)(2.5,10.)(0.,){/Straight}{1}
\FALabel(2.15636,7.28682)[tl]{$e$}
\FAProp(20.,17.)(16.,15.)(0.,){/Straight}{1}
\FALabel(17.7868,16.9064)[br]{$t$}
\FAProp(20.,10.)(12.,6.5)(0.,){/Straight}{-1}
\FALabel(14.6702,6.55096)[tl]{$t$}
\FAProp(20.,3.)(16.,15.)(0.,){/ScalarDash}{0}
\FALabel(17.87,12.55)[l]{$h^0$}
\FAProp(2.5,10.)(6.,10.)(0.,){/Sine}{0}
\FALabel(4.25,11.07)[b]{$Z$}
\FAProp(16.,15.)(12.,13.5)(0.,){/Straight}{1}
\FALabel(13.4558,15.2213)[b]{$t$}
\FAProp(12.,6.5)(6.,10.)(0.,){/ScalarDash}{-1}
\FALabel(8.699,7.39114)[tr]{$\tilde t^s$}
\FAProp(12.,6.5)(12.,13.5)(0.,){/Straight}{0}
\FALabel(12.82,10.)[l]{$\tilde g$}
\FAProp(6.,10.)(12.,13.5)(0.,){/ScalarDash}{-1}
\FALabel(8.699,12.6089)[br]{$\tilde t^t$}
\FAVert(2.5,10.){0}
\FAVert(16.,15.){0}
\FAVert(12.,6.5){0}
\FAVert(6.,10.){0}
\FAVert(12.,13.5){0}

\FADiagram{(3)}
\FAProp(0.,15.)(4.5,10.)(0.,){/Straight}{-1}
\FALabel(1.57789,11.9431)[tr]{$e$}
\FAProp(0.,5.)(4.5,10.)(0.,){/Straight}{1}
\FALabel(2.92211,6.9431)[tl]{$e$}
\FAProp(20.,17.)(13.5,16.)(0.,){/Straight}{1}
\FALabel(16.5143,17.552)[b]{$t$}
\FAProp(20.,10.)(16.,6.5)(0.,){/Straight}{-1}
\FALabel(18.5407,7.56351)[tl]{$t$}
\FAProp(20.,3.)(16.,6.5)(0.,){/ScalarDash}{0}
\FALabel(17.6239,4.25165)[tr]{$h^0$}
\FAProp(4.5,10.)(7.5,12.5)(0.,){/Sine}{0}
\FALabel(5.48771,11.9607)[br]{$\gamma$}
\FAProp(16.,6.5)(13.5,9.)(0.,){/Straight}{-1}
\FALabel(14.134,7.13398)[tr]{$t$}
\FAProp(13.5,16.)(7.5,12.5)(0.,){/ScalarDash}{1}
\FALabel(10.199,15.1089)[br]{$\tilde t^s$}
\FAProp(13.5,16.)(13.5,9.)(0.,){/Straight}{0}
\FALabel(14.32,12.5)[l]{$\tilde g$}
\FAProp(7.5,12.5)(13.5,9.)(0.,){/ScalarDash}{1}
\FALabel(10.199,9.89114)[tr]{$\tilde t^s$}
\FAVert(4.5,10.){0}
\FAVert(13.5,16.){0}
\FAVert(16.,6.5){0}
\FAVert(7.5,12.5){0}
\FAVert(13.5,9.){0}

\FADiagram{(4)}
\FAProp(0.,15.)(4.5,10.)(0.,){/Straight}{-1}
\FALabel(1.57789,11.9431)[tr]{$e$}
\FAProp(0.,5.)(4.5,10.)(0.,){/Straight}{1}
\FALabel(2.92211,6.9431)[tl]{$e$}
\FAProp(20.,17.)(13.5,16.)(0.,){/Straight}{1}
\FALabel(16.5143,17.552)[b]{$t$}
\FAProp(20.,10.)(16.,6.5)(0.,){/Straight}{-1}
\FALabel(18.5407,7.56351)[tl]{$t$}
\FAProp(20.,3.)(16.,6.5)(0.,){/ScalarDash}{0}
\FALabel(17.6239,4.25165)[tr]{$h^0$}
\FAProp(4.5,10.)(7.5,12.5)(0.,){/Sine}{0}
\FALabel(5.48771,11.9607)[br]{$Z$}
\FAProp(16.,6.5)(13.5,9.)(0.,){/Straight}{-1}
\FALabel(14.134,7.13398)[tr]{$t$}
\FAProp(13.5,16.)(7.5,12.5)(0.,){/ScalarDash}{1}
\FALabel(10.199,15.1089)[br]{$\tilde t^s$}
\FAProp(13.5,16.)(13.5,9.)(0.,){/Straight}{0}
\FALabel(14.32,12.5)[l]{$\tilde g$}
\FAProp(7.5,12.5)(13.5,9.)(0.,){/ScalarDash}{1}
\FALabel(10.199,9.89114)[tr]{$\tilde t^t$}
\FAVert(4.5,10.){0}
\FAVert(13.5,16.){0}
\FAVert(16.,6.5){0}
\FAVert(7.5,12.5){0}
\FAVert(13.5,9.){0}

\FADiagram{(5)}
\FAProp(0.,15.)(3.,10.)(0.,){/Straight}{-1}
\FALabel(0.650886,12.1825)[tr]{$e$}
\FAProp(0.,5.)(3.,10.)(0.,){/Straight}{1}
\FALabel(2.34911,7.18253)[tl]{$e$}
\FAProp(20.,17.)(8.5,12.5)(0.,){/Straight}{1}
\FALabel(13.6852,15.7134)[b]{$t$}
\FAProp(20.,10.)(16.,10.)(0.,){/Straight}{-1}
\FALabel(18.,11.07)[b]{$t$}
\FAProp(20.,3.)(16.,3.)(0.,){/ScalarDash}{0}
\FALabel(18.,2.18)[t]{$h^0$}
\FAProp(3.,10.)(8.5,12.5)(0.,){/Sine}{0}
\FALabel(5.58861,12.1811)[br]{$\gamma$}
\FAProp(8.5,12.5)(10.,6.5)(0.,){/Straight}{1}
\FALabel(8.22628,9.12407)[r]{$t$}
\FAProp(16.,10.)(16.,3.)(0.,){/ScalarDash}{-1}
\FALabel(17.07,6.5)[l]{$\tilde t^s$}
\FAProp(16.,10.)(10.,6.5)(0.,){/Straight}{0}
\FALabel(12.825,8.89291)[br]{$\tilde g$}
\FAProp(16.,3.)(10.,6.5)(0.,){/ScalarDash}{-1}
\FALabel(12.699,3.89114)[tr]{$\tilde t^t$}
\FAVert(3.,10.){0}
\FAVert(8.5,12.5){0}
\FAVert(16.,10.){0}
\FAVert(16.,3.){0}
\FAVert(10.,6.5){0}

\FADiagram{(6)}
\FAProp(0.,15.)(3.,10.)(0.,){/Straight}{-1}
\FALabel(0.650886,12.1825)[tr]{$e$}
\FAProp(0.,5.)(3.,10.)(0.,){/Straight}{1}
\FALabel(2.34911,7.18253)[tl]{$e$}
\FAProp(20.,17.)(8.5,12.5)(0.,){/Straight}{1}
\FALabel(13.6852,15.7134)[b]{$t$}
\FAProp(20.,10.)(16.,10.)(0.,){/Straight}{-1}
\FALabel(18.,11.07)[b]{$t$}
\FAProp(20.,3.)(16.,3.)(0.,){/ScalarDash}{0}
\FALabel(18.,2.18)[t]{$h^0$}
\FAProp(3.,10.)(8.5,12.5)(0.,){/Sine}{0}
\FALabel(5.58861,12.1811)[br]{$Z$}
\FAProp(8.5,12.5)(10.,6.5)(0.,){/Straight}{1}
\FALabel(8.22628,9.12407)[r]{$t$}
\FAProp(16.,10.)(16.,3.)(0.,){/ScalarDash}{-1}
\FALabel(17.07,6.5)[l]{$\tilde t^s$}
\FAProp(16.,10.)(10.,6.5)(0.,){/Straight}{0}
\FALabel(12.825,8.89291)[br]{$\tilde g$}
\FAProp(16.,3.)(10.,6.5)(0.,){/ScalarDash}{-1}
\FALabel(12.699,3.89114)[tr]{$\tilde t^t$}
\FAVert(3.,10.){0}
\FAVert(8.5,12.5){0}
\FAVert(16.,10.){0}
\FAVert(16.,3.){0}
\FAVert(10.,6.5){0}

\FADiagram{(7)}
\FAProp(0.,15.)(3.,9.)(0.,){/Straight}{-1}
\FALabel(2.09361,12.9818)[bl]{$e$}
\FAProp(0.,5.)(3.,9.)(0.,){/Straight}{1}
\FALabel(0.74,7.45)[br]{$e$}
\FAProp(20.,17.)(15.,17.)(0.,){/Straight}{1}
\FALabel(17.5,18.07)[b]{$t$}
\FAProp(20.,10.)(7.5,7.)(0.,){/Straight}{-1}
\FALabel(11.9083,6.8369)[t]{$t$}
\FAProp(20.,3.)(15.,10.)(0.,){/ScalarDash}{0}
\FALabel(17.93,4.73)[tr]{$h^0$}
\FAProp(3.,9.)(7.5,7.)(0.,){/Sine}{0}
\FALabel(5.10049,7.06359)[tr]{$\gamma$}
\FAProp(7.5,7.)(9.,13.5)(0.,){/Straight}{-1}
\FALabel(7.21969,10.5985)[r]{$t$}
\FAProp(15.,17.)(15.,10.)(0.,){/ScalarDash}{1}
\FALabel(16.07,13.5)[l]{$\tilde t^s$}
\FAProp(15.,17.)(9.,13.5)(0.,){/Straight}{0}
\FALabel(11.825,15.8929)[br]{$\tilde g$}
\FAProp(15.,10.)(9.,13.5)(0.,){/ScalarDash}{1}
\FALabel(11.699,10.8911)[tr]{$\tilde t^t$}
\FAVert(3.,9.){0}
\FAVert(15.,17.){0}
\FAVert(7.5,7.){0}
\FAVert(15.,10.){0}
\FAVert(9.,13.5){0}

\FADiagram{(8)}
\FAProp(0.,15.)(3.,9.)(0.,){/Straight}{-1}
\FALabel(2.09361,12.9818)[bl]{$e$}
\FAProp(0.,5.)(3.,9.)(0.,){/Straight}{1}
\FALabel(0.74,7.45)[br]{$e$}
\FAProp(20.,17.)(15.,17.)(0.,){/Straight}{1}
\FALabel(17.5,18.07)[b]{$t$}
\FAProp(20.,10.)(7.5,7.)(0.,){/Straight}{-1}
\FALabel(11.9083,6.8369)[t]{$t$}
\FAProp(20.,3.)(15.,10.)(0.,){/ScalarDash}{0}
\FALabel(17.93,4.73)[tr]{$h^0$}
\FAProp(3.,9.)(7.5,7.)(0.,){/Sine}{0}
\FALabel(5.10049,7.06359)[tr]{$Z$}
\FAProp(7.5,7.)(9.,13.5)(0.,){/Straight}{-1}
\FALabel(7.21969,10.5985)[r]{$t$}
\FAProp(15.,17.)(15.,10.)(0.,){/ScalarDash}{1}
\FALabel(16.07,13.5)[l]{$\tilde t^s$}
\FAProp(15.,17.)(9.,13.5)(0.,){/Straight}{0}
\FALabel(11.825,15.8929)[br]{$\tilde g$}
\FAProp(15.,10.)(9.,13.5)(0.,){/ScalarDash}{1}
\FALabel(11.699,10.8911)[tr]{$\tilde t^t$}
\FAVert(3.,9.){0}
\FAVert(15.,17.){0}
\FAVert(7.5,7.){0}
\FAVert(15.,10.){0}
\FAVert(9.,13.5){0}

\FADiagram{(9)}
\FAProp(0.,15.)(3.,10.)(0.,){/Straight}{-1}
\FALabel(2.34911,12.8175)[bl]{$e$}
\FAProp(0.,5.)(3.,10.)(0.,){/Straight}{1}
\FALabel(0.650886,7.81747)[br]{$e$}
\FAProp(20.,17.)(15.5,17.)(0.,){/Straight}{1}
\FALabel(17.75,18.07)[b]{$t$}
\FAProp(20.,10.)(15.5,10.)(0.,){/Straight}{-1}
\FALabel(17.75,8.93)[t]{$t$}
\FAProp(20.,3.)(8.,10.)(0.,){/ScalarDash}{0}
\FALabel(13.825,5.85709)[tr]{$h^0$}
\FAProp(3.,10.)(8.,10.)(0.,){/Sine}{0}
\FALabel(5.5,8.93)[t]{$Z$}
\FAProp(8.,10.)(9.5,13.5)(0.,){/ScalarDash}{0}
\FALabel(8.03511,11.7821)[br]{$A^0$}
\FAProp(15.5,17.)(15.5,10.)(0.,){/Straight}{0}
\FALabel(16.32,13.5)[l]{$\tilde g$}
\FAProp(15.5,17.)(9.5,13.5)(0.,){/ScalarDash}{1}
\FALabel(12.199,16.1089)[br]{$\tilde t^s$}
\FAProp(15.5,10.)(9.5,13.5)(0.,){/ScalarDash}{-1}
\FALabel(12.199,11.8911)[tr]{$\tilde t^t$}
\FAVert(3.,10.){0}
\FAVert(15.5,17.){0}
\FAVert(15.5,10.){0}
\FAVert(8.,10.){0}
\FAVert(9.5,13.5){0}

\FADiagram{(10)}
\FAProp(0.,15.)(3.,10.)(0.,){/Straight}{-1}
\FALabel(2.34911,12.8175)[bl]{$e$}
\FAProp(0.,5.)(3.,10.)(0.,){/Straight}{1}
\FALabel(0.650886,7.81747)[br]{$e$}
\FAProp(20.,17.)(15.5,17.)(0.,){/Straight}{1}
\FALabel(17.75,18.07)[b]{$t$}
\FAProp(20.,10.)(15.5,10.)(0.,){/Straight}{-1}
\FALabel(17.75,8.93)[t]{$t$}
\FAProp(20.,3.)(8.,10.)(0.,){/ScalarDash}{0}
\FALabel(13.825,5.85709)[tr]{$h^0$}
\FAProp(3.,10.)(8.,10.)(0.,){/Sine}{0}
\FALabel(5.5,8.93)[t]{$Z$}
\FAProp(8.,10.)(9.5,13.5)(0.,){/ScalarDash}{0}
\FALabel(8.03511,11.7821)[br]{$G^0$}
\FAProp(15.5,17.)(15.5,10.)(0.,){/Straight}{0}
\FALabel(16.32,13.5)[l]{$\tilde g$}
\FAProp(15.5,17.)(9.5,13.5)(0.,){/ScalarDash}{1}
\FALabel(12.199,16.1089)[br]{$\tilde t^s$}
\FAProp(15.5,10.)(9.5,13.5)(0.,){/ScalarDash}{-1}
\FALabel(12.199,11.8911)[tr]{$\tilde t^t$}
\FAVert(3.,10.){0}
\FAVert(15.5,17.){0}
\FAVert(15.5,10.){0}
\FAVert(8.,10.){0}
\FAVert(9.5,13.5){0}

\FADiagram{(11)}
\FAProp(0.,15.)(3.,10.)(0.,){/Straight}{-1}
\FALabel(2.34911,12.8175)[bl]{$e$}
\FAProp(0.,5.)(3.,10.)(0.,){/Straight}{1}
\FALabel(0.650886,7.81747)[br]{$e$}
\FAProp(20.,17.)(15.5,17.)(0.,){/Straight}{1}
\FALabel(17.75,18.07)[b]{$t$}
\FAProp(20.,10.)(15.5,10.)(0.,){/Straight}{-1}
\FALabel(17.75,8.93)[t]{$t$}
\FAProp(20.,3.)(8.,10.)(0.,){/ScalarDash}{0}
\FALabel(13.825,5.85709)[tr]{$h^0$}
\FAProp(3.,10.)(8.,10.)(0.,){/Sine}{0}
\FALabel(5.5,8.93)[t]{$Z$}
\FAProp(8.,10.)(9.5,13.5)(0.,){/Sine}{0}
\FALabel(7.80533,11.8806)[br]{$Z$}
\FAProp(15.5,17.)(15.5,10.)(0.,){/Straight}{0}
\FALabel(16.32,13.5)[l]{$\tilde g$}
\FAProp(15.5,17.)(9.5,13.5)(0.,){/ScalarDash}{1}
\FALabel(12.199,16.1089)[br]{$\tilde t^s$}
\FAProp(15.5,10.)(9.5,13.5)(0.,){/ScalarDash}{-1}
\FALabel(12.199,11.8911)[tr]{$\tilde t^t$}
\FAVert(3.,10.){0}
\FAVert(15.5,17.){0}
\FAVert(15.5,10.){0}
\FAVert(8.,10.){0}
\FAVert(9.5,13.5){0}

\FADiagram{(12)}
\FAProp(0.,15.)(3.,10.)(0.,){/Straight}{-1}
\FALabel(0.650886,12.1825)[tr]{$e$}
\FAProp(0.,5.)(3.,10.)(0.,){/Straight}{1}
\FALabel(2.34911,7.18253)[tl]{$e$}
\FAProp(20.,17.)(13.5,7.5)(0.,){/Straight}{1}
\FALabel(18.0237,15.8612)[br]{$t$}
\FAProp(20.,10.)(13.5,14.5)(0.,){/Straight}{-1}
\FALabel(18.5877,9.82445)[tr]{$t$}
\FAProp(20.,3.)(6.5,7.5)(0.,){/ScalarDash}{0}
\FALabel(12.8389,4.49671)[t]{$h^0$}
\FAProp(3.,10.)(6.5,14.5)(0.,){/Sine}{0}
\FALabel(4.0065,12.7216)[br]{$\gamma$}
\FAProp(13.5,7.5)(13.5,14.5)(0.,){/Straight}{0}
\FALabel(12.68,11.)[r]{$\tilde g$}
\FAProp(13.5,7.5)(6.5,7.5)(0.,){/ScalarDash}{1}
\FALabel(10.,8.57)[b]{$\tilde t^s$}
\FAProp(13.5,14.5)(6.5,14.5)(0.,){/ScalarDash}{-1}
\FALabel(10.,15.57)[b]{$\tilde t^t$}
\FAProp(6.5,7.5)(6.5,14.5)(0.,){/ScalarDash}{1}
\FALabel(6.43,9.3)[r]{$\tilde t^t$}
\FAVert(3.,10.){0}
\FAVert(13.5,7.5){0}
\FAVert(13.5,14.5){0}
\FAVert(6.5,7.5){0}
\FAVert(6.5,14.5){0}

\FADiagram{(13)}
\FAProp(0.,15.)(3.,10.)(0.,){/Straight}{-1}
\FALabel(0.650886,12.1825)[tr]{$e$}
\FAProp(0.,5.)(3.,10.)(0.,){/Straight}{1}
\FALabel(2.34911,7.18253)[tl]{$e$}
\FAProp(20.,17.)(13.5,7.5)(0.,){/Straight}{1}
\FALabel(18.0237,15.8612)[br]{$t$}
\FAProp(20.,10.)(13.5,14.5)(0.,){/Straight}{-1}
\FALabel(18.5877,9.82445)[tr]{$t$}
\FAProp(20.,3.)(6.5,7.5)(0.,){/ScalarDash}{0}
\FALabel(12.8389,4.49671)[t]{$h^0$}
\FAProp(3.,10.)(6.5,14.5)(0.,){/Sine}{0}
\FALabel(4.0065,12.7216)[br]{$Z$}
\FAProp(13.5,7.5)(13.5,14.5)(0.,){/Straight}{0}
\FALabel(12.68,11.)[r]{$\tilde g$}
\FAProp(13.5,7.5)(6.5,7.5)(0.,){/ScalarDash}{1}
\FALabel(10.,8.57)[b]{$\tilde t^s$}
\FAProp(13.5,14.5)(6.5,14.5)(0.,){/ScalarDash}{-1}
\FALabel(10.,15.57)[b]{$\tilde t^t$}
\FAProp(6.5,7.5)(6.5,14.5)(0.,){/ScalarDash}{1}
\FALabel(6.43,9.3)[r]{$\tilde t^u$}
\FAVert(3.,10.){0}
\FAVert(13.5,7.5){0}
\FAVert(13.5,14.5){0}
\FAVert(6.5,7.5){0}
\FAVert(6.5,14.5){0}

\FADiagram{(14)}
\FAProp(0.,15.)(2.5,10.)(0.,){/Straight}{-1}
\FALabel(0.343638,12.2868)[tr]{$e$}
\FAProp(0.,5.)(2.5,10.)(0.,){/Straight}{1}
\FALabel(2.15636,7.28682)[tl]{$e$}
\FAProp(20.,17.)(13.5,13.5)(0.,){/Straight}{1}
\FALabel(16.4951,16.1347)[br]{$t$}
\FAProp(20.,10.)(13.5,6.5)(0.,){/Straight}{-1}
\FALabel(17.0049,7.36527)[tl]{$t$}
\FAProp(20.,3.)(6.5,6.5)(0.,){/ScalarDash}{0}
\FALabel(12.9237,3.9716)[t]{$h^0$}
\FAProp(2.5,10.)(6.5,13.5)(0.,){/Sine}{0}
\FALabel(3.95932,12.4365)[br]{$\gamma$}
\FAProp(13.5,13.5)(13.5,6.5)(0.,){/Straight}{0}
\FALabel(14.47,10.95)[l]{$\tilde g$}
\FAProp(13.5,13.5)(6.5,13.5)(0.,){/ScalarDash}{1}
\FALabel(10.,14.57)[b]{$\tilde t^s$}
\FAProp(13.5,6.5)(6.5,6.5)(0.,){/ScalarDash}{-1}
\FALabel(10.,7.57)[b]{$\tilde t^t$}
\FAProp(6.5,6.5)(6.5,13.5)(0.,){/ScalarDash}{-1}
\FALabel(6.43,9.)[r]{$\tilde t^s$}
\FAVert(2.5,10.){0}
\FAVert(13.5,13.5){0}
\FAVert(13.5,6.5){0}
\FAVert(6.5,6.5){0}
\FAVert(6.5,13.5){0}

\FADiagram{(15)}
\FAProp(0.,15.)(2.5,10.)(0.,){/Straight}{-1}
\FALabel(0.343638,12.2868)[tr]{$e$}
\FAProp(0.,5.)(2.5,10.)(0.,){/Straight}{1}
\FALabel(2.15636,7.28682)[tl]{$e$}
\FAProp(20.,17.)(13.5,13.5)(0.,){/Straight}{1}
\FALabel(16.4951,16.1347)[br]{$t$}
\FAProp(20.,10.)(13.5,6.5)(0.,){/Straight}{-1}
\FALabel(17.0049,7.36527)[tl]{$t$}
\FAProp(20.,3.)(6.5,6.5)(0.,){/ScalarDash}{0}
\FALabel(12.9237,3.9716)[t]{$h^0$}
\FAProp(2.5,10.)(6.5,13.5)(0.,){/Sine}{0}
\FALabel(3.95932,12.4365)[br]{$Z$}
\FAProp(13.5,13.5)(13.5,6.5)(0.,){/Straight}{0}
\FALabel(14.47,10.95)[l]{$\tilde g$}
\FAProp(13.5,13.5)(6.5,13.5)(0.,){/ScalarDash}{1}
\FALabel(10.,14.57)[b]{$\tilde t^s$}
\FAProp(13.5,6.5)(6.5,6.5)(0.,){/ScalarDash}{-1}
\FALabel(10.,7.57)[b]{$\tilde t^t$}
\FAProp(6.5,6.5)(6.5,13.5)(0.,){/ScalarDash}{-1}
\FALabel(6.43,9.)[r]{$\tilde t^u$}
\FAVert(2.5,10.){0}
\FAVert(13.5,13.5){0}
\FAVert(13.5,6.5){0}
\FAVert(6.5,6.5){0}
\FAVert(6.5,13.5){0}

\FADiagram{(16)}
\FAProp(0.,15.)(3.,9.)(0.,){/Straight}{-1}
\FALabel(2.40636,12.2132)[bl]{$e$}
\FAProp(0.,5.)(3.,9.)(0.,){/Straight}{1}
\FALabel(0.74,7.45)[br]{$e$}
\FAProp(20.,17.)(13.5,16.5)(0.,){/Straight}{1}
\FALabel(16.6311,17.8154)[b]{$t$}
\FAProp(20.,10.)(6.,7.)(0.,){/Straight}{-1}
\FALabel(10.9524,6.83555)[t]{$t$}
\FAProp(20.,3.)(13.5,16.5)(0.,){/ScalarDash}{0}
\FALabel(16.42,12.47)[bl]{$h^0$}
\FAProp(3.,9.)(6.,7.)(0.,){/Sine}{0}
\FALabel(4.12021,7.19032)[tr]{$\gamma$}
\FAProp(13.5,16.5)(11.45,14.1)(0.,){/Straight}{1}
\FALabel(11.7764,15.8267)[br]{$t$}
\FAProp(6.,7.)(8.4,10.)(0.,){/Straight}{-1}
\FALabel(6.46965,8.98828)[br]{$t$}
\FAProp(11.45,14.1)(8.4,10.)(-0.8,){/Straight}{0}
\FALabel(12.128,12.5341)[tl]{$\tilde g$}
\FAProp(11.45,14.1)(8.4,10.)(0.8,){/ScalarDash}{1}
\FALabel(7.52137,13.7151)[br]{$\tilde t^s$}
\FAVert(3.,9.){0}
\FAVert(13.5,16.5){0}
\FAVert(6.,7.){0}
\FAVert(11.45,14.1){0}
\FAVert(8.4,10.){0}

\FADiagram{ (17)}
\FAProp(0.,15.)(3.,9.)(0.,){/Straight}{-1}
\FALabel(2.40636,12.2132)[bl]{$e$}
\FAProp(0.,5.)(3.,9.)(0.,){/Straight}{1}
\FALabel(0.74,7.45)[br]{$e$}
\FAProp(20.,17.)(13.5,16.5)(0.,){/Straight}{1}
\FALabel(16.6311,17.8154)[b]{$t$}
\FAProp(20.,10.)(6.,7.)(0.,){/Straight}{-1}
\FALabel(10.9524,6.83555)[t]{$t$}
\FAProp(20.,3.)(13.5,16.5)(0.,){/ScalarDash}{0}
\FALabel(16.42,12.47)[bl]{$h^0$}
\FAProp(3.,9.)(6.,7.)(0.,){/Sine}{0}
\FALabel(4.12021,7.19032)[tr]{$Z$}
\FAProp(13.5,16.5)(11.45,14.1)(0.,){/Straight}{1}
\FALabel(11.7764,15.8267)[br]{$t$}
\FAProp(6.,7.)(8.4,10.)(0.,){/Straight}{-1}
\FALabel(6.46965,8.98828)[br]{$t$}
\FAProp(11.45,14.1)(8.4,10.)(-0.8,){/Straight}{0}
\FALabel(12.128,12.5341)[tl]{$\tilde g$}
\FAProp(11.45,14.1)(8.4,10.)(0.8,){/ScalarDash}{1}
\FALabel(7.52137,13.7151)[br]{$\tilde t^s$}
\FAVert(3.,9.){0}
\FAVert(13.5,16.5){0}
\FAVert(6.,7.){0}
\FAVert(11.45,14.1){0}
\FAVert(8.4,10.){0}

\FADiagram{(18)}
\FAProp(0.,15.)(3.5,12.5)(0.,){/Straight}{-1}
\FALabel(1.32908,12.9687)[tr]{$e$}
\FAProp(0.,5.)(3.5,12.5)(0.,){/Straight}{1}
\FALabel(2.67458,8.57453)[tl]{$e$}
\FAProp(20.,17.)(8.,15.)(0.,){/Straight}{1}
\FALabel(13.7452,17.0489)[b]{$t$}
\FAProp(20.,10.)(14.5,6.)(0.,){/Straight}{-1}
\FALabel(17.6817,7.22646)[tl]{$t$}
\FAProp(20.,3.)(14.5,6.)(0.,){/ScalarDash}{0}
\FALabel(17.1075,3.83874)[tr]{$h^0$}
\FAProp(3.5,12.5)(8.,15.)(0.,){/Sine}{0}
\FALabel(5.47725,14.6249)[br]{$\gamma$}
\FAProp(8.,15.)(9.8,12.55)(0.,){/Straight}{1}
\FALabel(8.13088,13.3373)[tr]{$t$}
\FAProp(14.5,6.)(12.7,8.4)(0.,){/Straight}{-1}
\FALabel(12.84,6.75)[tr]{$t$}
\FAProp(9.8,12.55)(12.7,8.4)(0.8,){/Straight}{0}
\FALabel(9.0044,9.05036)[tr]{$\tilde g$}
\FAProp(9.8,12.55)(12.7,8.4)(-0.8,){/ScalarDash}{1}
\FALabel(13.7005,12.0428)[bl]{$\tilde t^s$}
\FAVert(3.5,12.5){0}
\FAVert(8.,15.){0}
\FAVert(14.5,6.){0}
\FAVert(9.8,12.55){0}
\FAVert(12.7,8.4){0}

\FADiagram{(19)}
\FAProp(0.,15.)(3.5,12.5)(0.,){/Straight}{-1}
\FALabel(1.32908,12.9687)[tr]{$e$}
\FAProp(0.,5.)(3.5,12.5)(0.,){/Straight}{1}
\FALabel(2.67458,8.57453)[tl]{$e$}
\FAProp(20.,17.)(8.,15.)(0.,){/Straight}{1}
\FALabel(13.7452,17.0489)[b]{$t$}
\FAProp(20.,10.)(14.5,6.)(0.,){/Straight}{-1}
\FALabel(17.6817,7.22646)[tl]{$t$}
\FAProp(20.,3.)(14.5,6.)(0.,){/ScalarDash}{0}
\FALabel(17.1075,3.83874)[tr]{$h^0$}
\FAProp(3.5,12.5)(8.,15.)(0.,){/Sine}{0}
\FALabel(5.47725,14.6249)[br]{$Z$}
\FAProp(8.,15.)(9.8,12.55)(0.,){/Straight}{1}
\FALabel(8.13088,13.3373)[tr]{$t$}
\FAProp(14.5,6.)(12.7,8.4)(0.,){/Straight}{-1}
\FALabel(12.84,6.75)[tr]{$t$}
\FAProp(9.8,12.55)(12.7,8.4)(0.8,){/Straight}{0}
\FALabel(9.0044,9.05036)[tr]{$\tilde g$}
\FAProp(9.8,12.55)(12.7,8.4)(-0.8,){/ScalarDash}{1}
\FALabel(13.7005,12.0428)[bl]{$\tilde t^s$}
\FAVert(3.5,12.5){0}
\FAVert(8.,15.){0}
\FAVert(14.5,6.){0}
\FAVert(9.8,12.55){0}
\FAVert(12.7,8.4){0}

\end{feynartspicture}

\caption[]{ Virtual diagrams of $e^+e^- \rightarrow t \bar t h^0$, where $s, t, u$ are
 indexes of squark and solid lines in loop represent gluino.}
\end{figure}

\begin{figure}

\unitlength=1bp%

\begin{feynartspicture}(432,504)(4,5.3)

\FADiagram{(1)}
\FAProp(0.,15.)(4.5,10.)(0.,){/Straight}{-1}
\FALabel(1.57789,11.9431)[tr]{$e$}
\FAProp(0.,5.)(4.5,10.)(0.,){/Straight}{1}
\FALabel(2.92211,6.9431)[tl]{$e$}
\FAProp(20.,17.)(13.,14.5)(0.,){/Straight}{1}
\FALabel(15.9787,16.7297)[b]{$t$}
\FAProp(20.,10.)(10.85,8.4)(0.,){/Straight}{-1}
\FALabel(18.4569,10.7663)[b]{$t$}
\FAProp(20.,3.)(13.,14.5)(0.,){/ScalarDash}{0}
\FALabel(17.98,4.93)[tr]{$h^0$}
\FAProp(4.5,10.)(10.85,8.4)(0.,){/Sine}{0}
\FALabel(7.29629,8.17698)[t]{$\gamma$}
\FAProp(13.,14.5)(10.85,8.4)(0.,){/Straight}{1}
\FALabel(10.9431,11.9652)[r]{$t$}
\FAVert(4.5,10.){0}
\FAVert(13.,14.5){0}
\FAVert(10.85,8.4){1}

\FADiagram{(2)}
\FAProp(0.,15.)(4.5,10.)(0.,){/Straight}{-1}
\FALabel(1.57789,11.9431)[tr]{$e$}
\FAProp(0.,5.)(4.5,10.)(0.,){/Straight}{1}
\FALabel(2.92211,6.9431)[tl]{$e$}
\FAProp(20.,17.)(13.,14.5)(0.,){/Straight}{1}
\FALabel(15.9787,16.7297)[b]{$t$}
\FAProp(20.,10.)(10.85,8.4)(0.,){/Straight}{-1}
\FALabel(18.4569,10.7663)[b]{$t$}
\FAProp(20.,3.)(13.,14.5)(0.,){/ScalarDash}{0}
\FALabel(17.98,4.93)[tr]{$h^0$}
\FAProp(4.5,10.)(10.85,8.4)(0.,){/Sine}{0}
\FALabel(7.29629,8.17698)[t]{$Z$}
\FAProp(13.,14.5)(10.85,8.4)(0.,){/Straight}{1}
\FALabel(10.9431,11.9652)[r]{$t$}
\FAVert(4.5,10.){0}
\FAVert(13.,14.5){0}
\FAVert(10.85,8.4){1}

\FADiagram{(3)}
\FAProp(0.,15.)(5.5,10.)(0.,){/Straight}{-1}
\FALabel(2.18736,11.8331)[tr]{$e$}
\FAProp(0.,5.)(5.5,10.)(0.,){/Straight}{1}
\FALabel(3.31264,6.83309)[tl]{$e$}
\FAProp(20.,17.)(11.5,10.)(0.,){/Straight}{1}
\FALabel(15.2447,14.2165)[br]{$t$}
\FAProp(20.,10.)(15.5,6.5)(0.,){/Straight}{-1}
\FALabel(17.2784,8.9935)[br]{$t$}
\FAProp(20.,3.)(15.5,6.5)(0.,){/ScalarDash}{0}
\FALabel(18.0681,5.29616)[bl]{$h^0$}
\FAProp(5.5,10.)(11.5,10.)(0.,){/Sine}{0}
\FALabel(8.5,11.07)[b]{$\gamma$}
\FAProp(11.5,10.)(15.5,6.5)(0.,){/Straight}{1}
\FALabel(12.9593,7.56351)[tr]{$t$}
\FAVert(5.5,10.){0}
\FAVert(11.5,10.){0}
\FAVert(15.5,6.5){1}

\FADiagram{(4)}
\FAProp(0.,15.)(5.5,10.)(0.,){/Straight}{-1}
\FALabel(2.18736,11.8331)[tr]{$e$}
\FAProp(0.,5.)(5.5,10.)(0.,){/Straight}{1}
\FALabel(3.31264,6.83309)[tl]{$e$}
\FAProp(20.,17.)(11.5,10.)(0.,){/Straight}{1}
\FALabel(15.2447,14.2165)[br]{$t$}
\FAProp(20.,10.)(15.5,6.5)(0.,){/Straight}{-1}
\FALabel(17.2784,8.9935)[br]{$t$}
\FAProp(20.,3.)(15.5,6.5)(0.,){/ScalarDash}{0}
\FALabel(18.0681,5.29616)[bl]{$h^0$}
\FAProp(5.5,10.)(11.5,10.)(0.,){/Sine}{0}
\FALabel(8.5,11.07)[b]{$Z$}
\FAProp(11.5,10.)(15.5,6.5)(0.,){/Straight}{1}
\FALabel(12.9593,7.56351)[tr]{$t$}
\FAVert(5.5,10.){0}
\FAVert(11.5,10.){0}
\FAVert(15.5,6.5){1}

\FADiagram{(5)}
\FAProp(0.,15.)(5.5,10.)(0.,){/Straight}{-1}
\FALabel(2.18736,11.8331)[tr]{$e$}
\FAProp(0.,5.)(5.5,10.)(0.,){/Straight}{1}
\FALabel(3.31264,6.83309)[tl]{$e$}
\FAProp(20.,17.)(11.5,10.)(0.,){/Straight}{1}
\FALabel(15.2447,14.2165)[br]{$t$}
\FAProp(20.,10.)(15.5,6.5)(0.,){/Straight}{-1}
\FALabel(17.2784,8.9935)[br]{$t$}
\FAProp(20.,3.)(15.5,6.5)(0.,){/ScalarDash}{0}
\FALabel(18.0681,5.29616)[bl]{$h^0$}
\FAProp(5.5,10.)(11.5,10.)(0.,){/Sine}{0}
\FALabel(8.5,11.07)[b]{$\gamma$}
\FAProp(11.5,10.)(15.5,6.5)(0.,){/Straight}{1}
\FALabel(12.9593,7.56351)[tr]{$t$}
\FAVert(5.5,10.){0}
\FAVert(15.5,6.5){0}
\FAVert(11.5,10.){1}

\FADiagram{(6)}
\FAProp(0.,15.)(5.5,10.)(0.,){/Straight}{-1}
\FALabel(2.18736,11.8331)[tr]{$e$}
\FAProp(0.,5.)(5.5,10.)(0.,){/Straight}{1}
\FALabel(3.31264,6.83309)[tl]{$e$}
\FAProp(20.,17.)(11.5,10.)(0.,){/Straight}{1}
\FALabel(15.2447,14.2165)[br]{$t$}
\FAProp(20.,10.)(15.5,6.5)(0.,){/Straight}{-1}
\FALabel(17.2784,8.9935)[br]{$t$}
\FAProp(20.,3.)(15.5,6.5)(0.,){/ScalarDash}{0}
\FALabel(18.0681,5.29616)[bl]{$h^0$}
\FAProp(5.5,10.)(11.5,10.)(0.,){/Sine}{0}
\FALabel(8.5,11.07)[b]{$Z$}
\FAProp(11.5,10.)(15.5,6.5)(0.,){/Straight}{1}
\FALabel(12.9593,7.56351)[tr]{$t$}
\FAVert(5.5,10.){0}
\FAVert(15.5,6.5){0}
\FAVert(11.5,10.){1}

\FADiagram{(7)}
\FAProp(0.,15.)(4.5,10.)(0.,){/Straight}{-1}
\FALabel(1.57789,11.9431)[tr]{$e$}
\FAProp(0.,5.)(4.5,10.)(0.,){/Straight}{1}
\FALabel(2.92211,6.9431)[tl]{$e$}
\FAProp(20.,17.)(13.,14.5)(0.,){/Straight}{1}
\FALabel(15.9787,16.7297)[b]{$t$}
\FAProp(20.,10.)(10.85,8.4)(0.,){/Straight}{-1}
\FALabel(18.4569,10.7663)[b]{$t$}
\FAProp(20.,3.)(13.,14.5)(0.,){/ScalarDash}{0}
\FALabel(17.98,4.93)[tr]{$h^0$}
\FAProp(4.5,10.)(10.85,8.4)(0.,){/Sine}{0}
\FALabel(7.29629,8.17698)[t]{$\gamma$}
\FAProp(13.,14.5)(10.85,8.4)(0.,){/Straight}{1}
\FALabel(10.9431,11.9652)[r]{$t$}
\FAVert(4.5,10.){0}
\FAVert(10.85,8.4){0}
\FAVert(13.,14.5){1}

\FADiagram{(8)}
\FAProp(0.,15.)(4.5,10.)(0.,){/Straight}{-1}
\FALabel(1.57789,11.9431)[tr]{$e$}
\FAProp(0.,5.)(4.5,10.)(0.,){/Straight}{1}
\FALabel(2.92211,6.9431)[tl]{$e$}
\FAProp(20.,17.)(13.,14.5)(0.,){/Straight}{1}
\FALabel(15.9787,16.7297)[b]{$t$}
\FAProp(20.,10.)(10.85,8.4)(0.,){/Straight}{-1}
\FALabel(18.4569,10.7663)[b]{$t$}
\FAProp(20.,3.)(13.,14.5)(0.,){/ScalarDash}{0}
\FALabel(17.98,4.93)[tr]{$h^0$}
\FAProp(4.5,10.)(10.85,8.4)(0.,){/Sine}{0}
\FALabel(7.29629,8.17698)[t]{$Z$}
\FAProp(13.,14.5)(10.85,8.4)(0.,){/Straight}{1}
\FALabel(10.9431,11.9652)[r]{$t$}
\FAVert(4.5,10.){0}
\FAVert(10.85,8.4){0}
\FAVert(13.,14.5){1}

\FADiagram{(9)}
\FAProp(0.,15.)(5.5,10.)(0.,){/Straight}{-1}
\FALabel(2.18736,11.8331)[tr]{$e$}
\FAProp(0.,5.)(5.5,10.)(0.,){/Straight}{1}
\FALabel(3.31264,6.83309)[tl]{$e$}
\FAProp(20.,17.)(15.5,13.5)(0.,){/Straight}{1}
\FALabel(17.2784,15.9935)[br]{$t$}
\FAProp(20.,10.)(15.5,13.5)(0.,){/Straight}{-1}
\FALabel(18.2216,12.4935)[bl]{$t$}
\FAProp(20.,3.)(12.,10.)(0.,){/ScalarDash}{0}
\FALabel(15.6239,6.00165)[tr]{$h^0$}
\FAProp(5.5,10.)(12.,10.)(0.,){/Sine}{0}
\FALabel(8.75,8.93)[t]{$Z$}
\FAProp(15.5,13.5)(12.,10.)(0.,){/ScalarDash}{0}
\FALabel(13.3108,12.1892)[br]{$G^0$}
\FAVert(5.5,10.){0}
\FAVert(12.,10.){0}
\FAVert(15.5,13.5){1}

\FADiagram{(10)}
\FAProp(0.,15.)(5.5,10.)(0.,){/Straight}{-1}
\FALabel(2.18736,11.8331)[tr]{$e$}
\FAProp(0.,5.)(5.5,10.)(0.,){/Straight}{1}
\FALabel(3.31264,6.83309)[tl]{$e$}
\FAProp(20.,17.)(15.5,13.5)(0.,){/Straight}{1}
\FALabel(17.2784,15.9935)[br]{$t$}
\FAProp(20.,10.)(15.5,13.5)(0.,){/Straight}{-1}
\FALabel(18.2216,12.4935)[bl]{$t$}
\FAProp(20.,3.)(12.,10.)(0.,){/ScalarDash}{0}
\FALabel(15.6239,6.00165)[tr]{$h^0$}
\FAProp(5.5,10.)(12.,10.)(0.,){/Sine}{0}
\FALabel(8.75,8.93)[t]{$Z$}
\FAProp(15.5,13.5)(12.,10.)(0.,){/Sine}{0}
\FALabel(13.134,12.366)[br]{$Z$}
\FAVert(5.5,10.){0}
\FAVert(12.,10.){0}
\FAVert(15.5,13.5){1}

\FADiagram{(11)}
\FAProp(0.,15.)(3.,10.)(0.,){/Straight}{-1}
\FALabel(0.650886,12.1825)[tr]{$e$}
\FAProp(0.,5.)(3.,10.)(0.,){/Straight}{1}
\FALabel(2.34911,7.18253)[tl]{$e$}
\FAProp(20.,17.)(15.5,5.5)(0.,){/Straight}{1}
\FALabel(19.2028,12.2089)[l]{$t$}
\FAProp(20.,10.)(8.,12.)(0.,){/Straight}{-1}
\FALabel(14.2548,12.0489)[b]{$t$}
\FAProp(20.,3.)(15.5,5.5)(0.,){/ScalarDash}{0}
\FALabel(17.5987,3.5936)[tr]{$h^0$}
\FAProp(12.,8.5)(15.5,5.5)(0.,){/Straight}{-1}
\FALabel(13.2213,6.30315)[tr]{$t$}
\FAProp(12.,8.5)(8.,12.)(0.,){/Straight}{1}
\FALabel(9.45932,9.56351)[tr]{$t$}
\FAProp(3.,10.)(8.,12.)(0.,){/Sine}{0}
\FALabel(4.92434,11.9591)[b]{$\gamma$}
\FAVert(3.,10.){0}
\FAVert(15.5,5.5){0}
\FAVert(8.,12.){0}
\FAVert(12.,8.5){1}

\FADiagram{(12)}
\FAProp(0.,15.)(3.,10.)(0.,){/Straight}{-1}
\FALabel(0.650886,12.1825)[tr]{$e$}
\FAProp(0.,5.)(3.,10.)(0.,){/Straight}{1}
\FALabel(2.34911,7.18253)[tl]{$e$}
\FAProp(20.,17.)(15.5,5.5)(0.,){/Straight}{1}
\FALabel(19.2028,12.2089)[l]{$t$}
\FAProp(20.,10.)(8.,12.)(0.,){/Straight}{-1}
\FALabel(14.2548,12.0489)[b]{$t$}
\FAProp(20.,3.)(15.5,5.5)(0.,){/ScalarDash}{0}
\FALabel(17.5987,3.5936)[tr]{$h^0$}
\FAProp(12.,8.5)(15.5,5.5)(0.,){/Straight}{-1}
\FALabel(13.2213,6.30315)[tr]{$t$}
\FAProp(12.,8.5)(8.,12.)(0.,){/Straight}{1}
\FALabel(9.45932,9.56351)[tr]{$t$}
\FAProp(3.,10.)(8.,12.)(0.,){/Sine}{0}
\FALabel(4.92434,11.9591)[b]{$Z$}
\FAVert(3.,10.){0}
\FAVert(15.5,5.5){0}
\FAVert(8.,12.){0}
\FAVert(12.,8.5){1}

\FADiagram{(13)}
\FAProp(0,15.)(6.,10.)(0,){/Straight}{-1}
\FALabel(2.48771,11.7893)[tr]{$e$}
\FAProp(0,5.)(6.,10.)(0,){/Straight}{1}
\FALabel(3.51229,6.78926)[tl]{$e$}
\FAProp(20.,17.)(13.,14.)(0,){/Straight}{1}
\FALabel(16.3694,16.4447)[br]{$t$}
\FAProp(20.,10.)(13.,7.)(0,){/Straight}{-1}
\FALabel(16.3694,9.44467)[br]{$t$}
\FAProp(20.,3.)(13.,7.)(0,){/ScalarDash}{0}
\FALabel(16.665,5.64872)[bl]{$h^0$}
\FAProp(13.,10.5)(13.,14.)(0,){/Straight}{-1}
\FALabel(14.07,12.25)[l]{$t$}
\FAProp(13.,10.5)(13.,7.)(0,){/Straight}{1}
\FALabel(11.93,8.75)[r]{$t$}
\FAProp(6.,10.)(13.,14.)(0,){/Sine}{0}
\FALabel(7.78902,14.1342)[tl]{$\gamma$}
\FAVert(6.,10.){0}
\FAVert(13.,14.){0}
\FAVert(13.,7.){0}
\FAVert(13.,10.5){1}

\FADiagram{(14)}
\FAProp(0,15.)(6.,10.)(0,){/Straight}{-1}
\FALabel(2.48771,11.7893)[tr]{$e$}
\FAProp(0,5.)(6.,10.)(0,){/Straight}{1}
\FALabel(3.51229,6.78926)[tl]{$e$}
\FAProp(20.,17.)(13.,14.)(0,){/Straight}{1}
\FALabel(16.3694,16.4447)[br]{$t$}
\FAProp(20.,10.)(13.,7.)(0,){/Straight}{-1}
\FALabel(16.3694,9.44467)[br]{$t$}
\FAProp(20.,3.)(13.,7.)(0,){/ScalarDash}{0}
\FALabel(16.665,5.64872)[bl]{$h^0$}
\FAProp(13.,10.5)(13.,14.)(0,){/Straight}{-1}
\FALabel(14.07,12.25)[l]{$t$}
\FAProp(13.,10.5)(13.,7.)(0,){/Straight}{1}
\FALabel(11.93,8.75)[r]{$t$}
\FAProp(6.,10.)(13.,14.)(0,){/Sine}{0}
\FALabel(7.78902,14.1342)[tl]{$Z$}
\FAVert(6.,10.){0}
\FAVert(13.,14.){0}
\FAVert(13.,7.){0}
\FAVert(13.,10.5){1}

\FADiagram{(15)}
\FAProp(0.,15.)(5.5,10.)(0.,){/Straight}{-1}
\FALabel(2.18736,11.8331)[tr]{$e$}
\FAProp(0.,5.)(5.5,10.)(0.,){/Straight}{1}
\FALabel(3.31264,6.83309)[tl]{$e$}
\FAProp(20.,17.)(15.5,13.5)(0.,){/Straight}{1}
\FALabel(17.2784,15.9935)[br]{$t$}
\FAProp(20.,10.)(15.5,13.5)(0.,){/Straight}{-1}
\FALabel(18.2216,12.4935)[bl]{$t$}
\FAProp(20.,3.)(12.,10.)(0.,){/ScalarDash}{0}
\FALabel(15.6239,6.00165)[tr]{$h^0$}
\FAProp(5.5,10.)(12.,10.)(0.,){/Sine}{0}
\FALabel(8.75,8.93)[t]{$Z$}
\FAProp(15.5,13.5)(12.,10.)(0.,){/ScalarDash}{0}
\FALabel(13.3108,12.1892)[br]{$A^0$}
\FAVert(5.5,10.){0}
\FAVert(12.,10.){0}
\FAVert(15.5,13.5){1}

\end{feynartspicture}

\caption[]{ Counter-term diagrams of $e^+e^- \rightarrow t \bar t h^0$.}
\end{figure}





\begin{figure}
\epsfxsize=15 cm \centerline{\epsffile{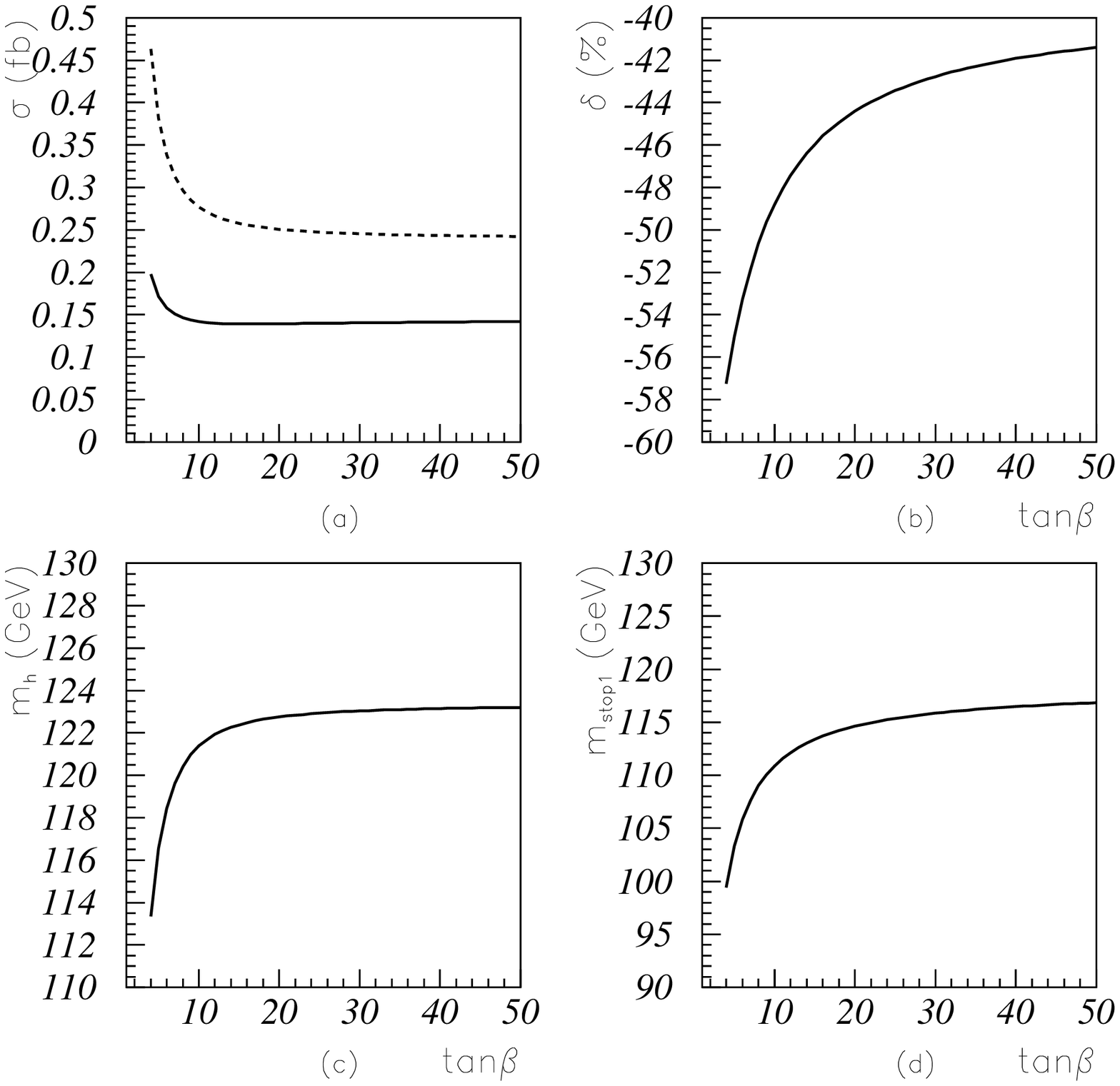}} \caption[]{ Cross
sections (a),  relative correction (b), Higgs mass (c) and lighter
stop mass (d) as a function of $\tan\beta$ for $\sqrt{s}=0.5$ TeV
with $A_t=1$ TeV, $M_{S}=400$ GeV, $\mu=-100$ GeV, $m_{A^0}=300$
GeV, $m_{\tilde{g}}=200$ GeV. The solid and dashed lines in (a)
represent the cross sections at NLO and LO.} \label{fig4}
\end{figure}

\begin{figure}
\epsfxsize=15 cm \centerline{\epsffile{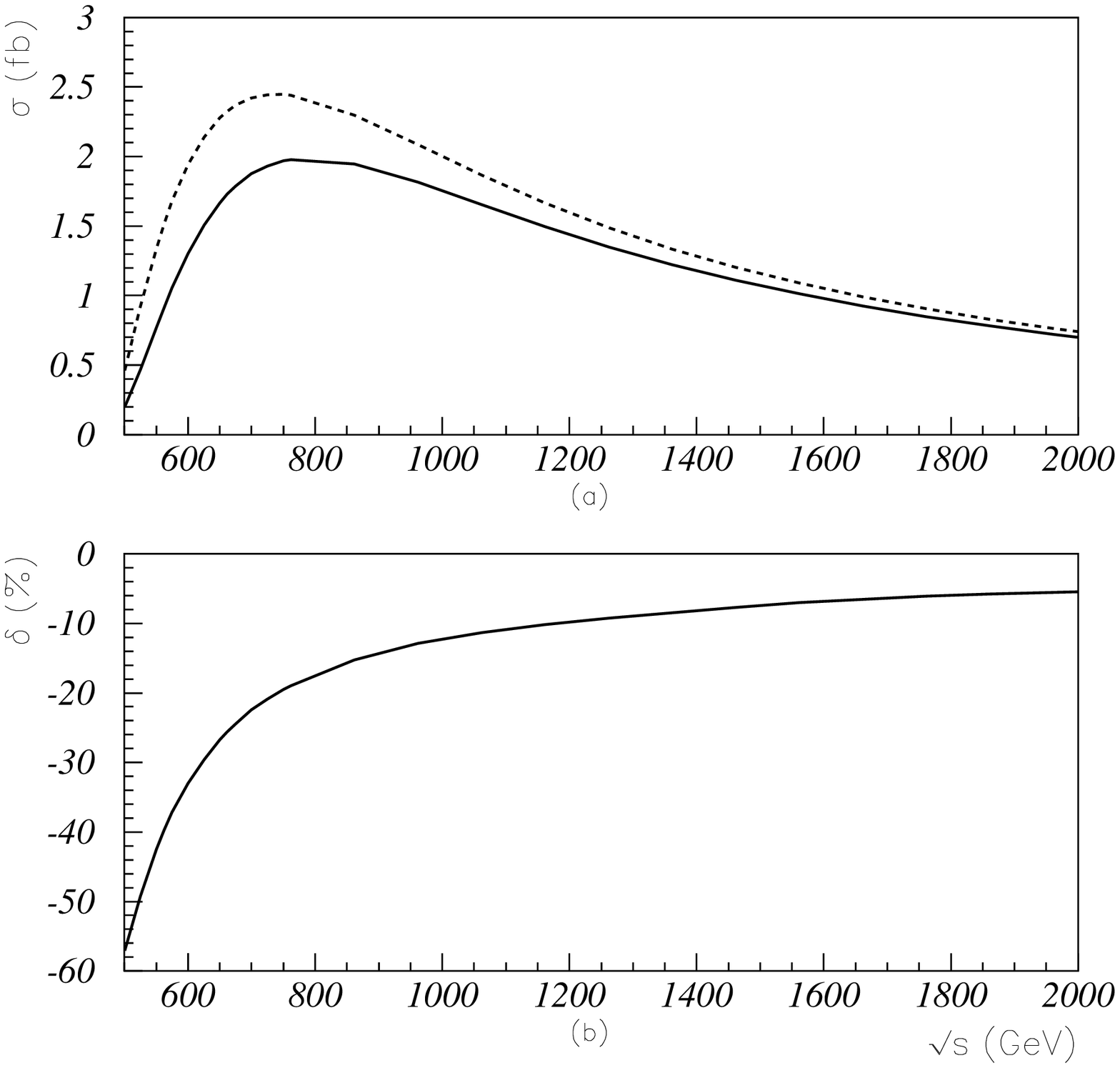}} \caption[]{ Cross
sections (a) and relative correction (b) as a function of
$\sqrt{s}$ for $\tan\beta=4$, where $m_{h^0}=113.3$ GeV and
$m_{\tilde{t}_1}= 99.4$ GeV. Other parameters and conventions are
the same with Fig. \ref{fig4}. }
\end{figure}

\begin{figure}
\epsfxsize=15 cm \centerline{\epsffile{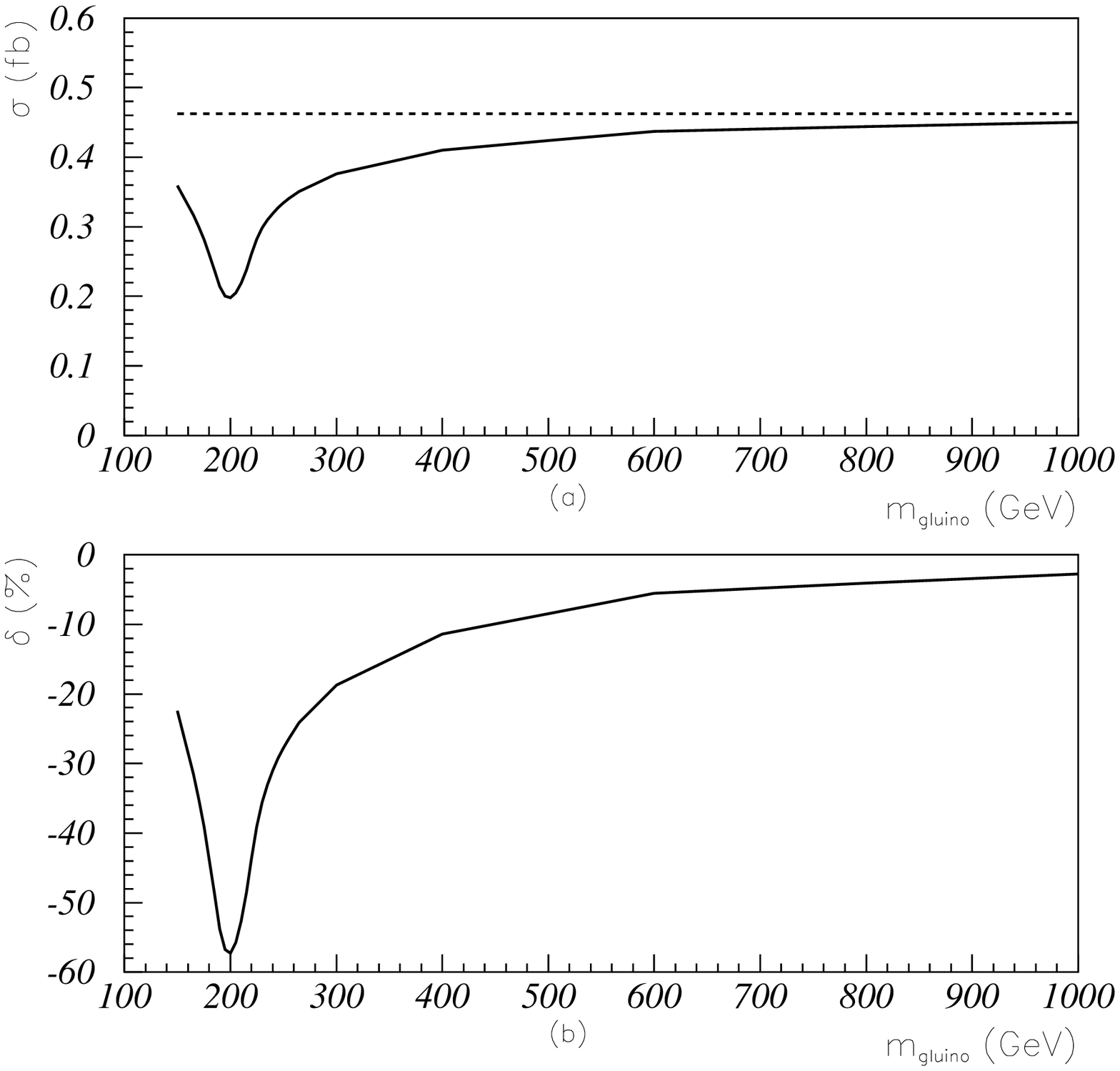}} \caption[]{ Cross
sections (a) and relative correction (b) as a function of
$m_{\tilde{g}} $ with $\tan\beta=4$, where $m_{h^0}=113.3$ GeV and
$m_{\tilde{t}_1}= 99.4$ GeV. Other parameters and conventions are
the same with Fig. \ref{fig4}. }
\end{figure}


\newpage

\begin{table}[t]
\centering
\begin{tabular}{|c|c|c|c|c|c|c|c|} \hline
$A_t$ (GeV)  &1000  &  950  &  900   & 800    & 700   & 500 & 200 \\
$m_{h^0}$ (GeV) & 113.4 & 115.2 & 116.1 & 115.7 &113.5 & 106.8 &
97.6 \\
$m_{\tilde{t}_1}$ (GeV) & 99.4 & 136.4 & 165.3 & 211.5 & 249.4 &
311.5 & 386.4 \\
LO           & 0.463 & 0.411  &  0.389 & 0.400  & 0.459 & 0.674 & 1.075\\
NLO          & 0.198 & 0.327  & 0.348  & 0.387  &0.452  & 0.670 & 1.077\\
$\delta$ (\%)& -57.2 & -20.4  & -10.4  & -3.2   & -1.6  & -0.7  & 0.2 \\
\hline
$A_t$ (GeV)  &0  &  -200  &  -500   & -700    & -800   & -900 & -1000 \\
$m_{h^0}$ (GeV) & 94.7 & 96.6 & 105.1 & 112.0 & 114.7 & 116.2 &
115.2 \\
$m_{\tilde{t}_1}$ (GeV) & 429.1 & 397.5 &325.2 & 266.3 & 231.2 &
189.8 & 136.4\\
LO           & 1.226 & 1.128  &  0.741 & 0.503  & 0.425 & 0.387 & 0.411\\
NLO          & 1.229 & 1.131  & 0.741  & 0.506  &0.432  & 0.393 & 0.410\\
$\delta$ (\%)& 0.3 & 0.3      & 0  & 0.5   & 1.7  &  1.6 & -0.2\\
\hline
\end{tabular}
\caption[]{Total cross sections [fb] at LO and NLO as well as
the relative corrections
as a function of $A_t$ with $\tan\beta=4$. Other parameters are the
same with Fig. \ref{fig4}.
}
\label{tb1}
\end{table}

\end{document}